# Balotage in Argentina 2015, a sentiment analysis of tweets

## *(Balotaje Argentina 2015 a partir de un análisis de sentimiento de tweets)*

Daniel Robins[12], Fernando Emmanuel Frati[1], Jonatan Alvarez[12], Jose Texier[13]

[1]Departamento de Ciencias Básicas y Tecnológicas, Universidad Nacional de Chilecito

9 de Julio 22, Chilecito, La Rioja, Argentina

daniel.robins@kunan.com.ar, fefrati@undec.edu.ar, jonatan.alvarez@kunan.com.ar, jtexier@undec.edu.ar

[2]KUNAN, Córdoba, Argentina

[3]Universidad Nacional Experimental del Táchira, Venezuela

ESTRUCTURA DEL ARTÍCULO

## Abstract

Twitter social network contains a large amount of information generated by its users. That information is composed of opinions and comments that may reflect trends in social behavior. There is talk of trend when it is possible to identify opinions and comments geared towards the same shared by a lot of people direction. To determine if two or more written opinions share the same address, techniques Natural Language Processing (NLP) are used. This paper proposes a methodology for predicting reflected in Twitter from the use of sentiment analysis functions NLP based on social behaviors. The case study was selected the 2015 Presidential in Argentina, and a software architecture Big Data composed Vertica data base with the component called Pulse was used. Through the analysis it was possible to detect trends in voting intentions with regard to the presidential candidates, achieving greater accuracy in predicting that achieved with traditional systems surveys.

**Resumen**

La red social Twitter contiene una gran cantidad de información generada por sus usuarios. Esa información está compuesta de opiniones y comentarios que pueden reflejar tendencias de comportamientos sociales. Se habla de tendencia cuando es posible identificar opiniones y comentarios orientadas hacia la misma dirección compartidas por una gran cantidad de personas. Para determinar si dos o más opiniones escritas comparten una misma dirección, se utilizan técnicas de Procesamiento de Lenguaje Natural (PLN). Este trabajo propone una metodología de predicción de comportamientos sociales reflejados en Twitter a partir del uso de funciones de análisis de sentimientos basadas en PLN. El caso de estudio seleccionado fue el Balotaje Presidencial de Argentina 2015, y se utilizó una arquitectura de software de Big Data compuesta por la base de Datos Vertica con el componente llamado Pulse. A través del análisis realizado fue posible detectar la tendencia en intención de voto con respecto a los candidatos presidenciales, consiguiendo mayor precisión en la predicción que la alcanzada con sistemas de encuestas tradicionales.

**Palabras clave:** PLN, Process Language Natural, Sentiment Analysis, Survey, Big Data, Ballotage, 2015, Argentina, Vertica, Pulse

## 1. Introducción

La necesidad de administrar grandes volúmenes de información no estructurada (heterogénea) es un tema que en la actualidad está en crecimiento (Canuto 2016). Evidencia de ello es el auge de los dispositivos móviles, sensores, aplicaciones de Internet, como así también empresas u organismos estatales, todos producen información digital. Con base en lo anterior, se denomina Big Data (BD) a la tecnología que permite procesar grandes volúmenes de datos en poco tiempo (Rajaraman, 2011). Por ello, el procesamiento y posterior análisis de los datos y/o información debe ser realizado preferiblemente en tiempo real para poder mejorar la toma de decisiones con base en la información generada. La caracterización típica de un problema de Big Data es si cumple con un gran volumen, con variedad de fuentes (tanto estructuradas o no), requiere velocidad y frecuencia de las actualizaciones y con veracidad de la información (tascon2013[1]).

Actualmente, Big Data se encuentra en un gran crecimiento debido a que las diversas áreas de las instituciones y/o empresas necesitan de los servicios que se pueden generar a partir de este tratamiento dinámico de la información, que por lo general se encuentra digital (Yang Yu 2015 / Sayan Unankard 2014 / Varsha Sahayak 2015). Por ello, el análisis de sentimientos (AS) a través de los datos de Twitter (tweets) bajo el enfoque de BD, cubre un nicho de investigación que está creciendo y que permite hacer propuestas metodológicas para problemas actuales y futuros. El AS ha sido estudiado desde principios de la década del 2000 con trabajos con enfoques basados en la polaridad ((Pang, Lee, y Vaithyanathan, 2002)Turney, 2002 / Cotelo 2013), ya que el AS es visto como sentencias o palabras que representan una opinión o perspectiva que puede ser evaluada como positiva, negativa, neutral o ilegible (Mike Thelwall 2011, 2012 / Agarwal 2011 / M. Ghiassi 2013). Igualmente, el hecho de poder encontrar las orientaciones semánticas sobre palabras o frases (tweets) en tiempo real a través de una red social de microblogging (Twitter), nos lleva a tomar consideraciones sobre la gramática y códigos propios de la red como los emoticonos (emoticons), etiquetas (hashtags), menciones (mentions), abreviaturas, jergas, entre otros (Agarwal 2011 / Varsha Sahayak 2015 / Johan Bollen 2011).

---

[1] Tascón, M. "Introducción: Big Data. Pasado, presente y futuro - Díalnet." 2013. <https://dialnet.unirioja.es/servlet/articulo?codigo=4423775>

La base del análisis de sentimientos de mensajes no estructurados, generalmente está centrada en el análisis semántico y por ende, en las técnicas clásicas de procesamiento de lenguaje natural (PLN). Es importante destacar que los tweets (además de cualquier otra información disponible) se pueden analizar también de las siguientes maneras: extracción de información basada en ontologías, minería de datos, clasificadores, técnicas de aprendizaje automático, entre otras(Cotelo 2013) [2]. Muchos trabajos continúan usando estas técnicas, combinaciones de ellas y generando nuevas propuestas sobre este dominio (Liu, 2012).

Esta época donde las relaciones sociales, culturales y económicas se encuentran fuertemente atravesadas por las Tecnologías de la Información y las Comunicaciones (TIC), ha dado lugar a las denominadas sociedades de la información y el conocimiento (Murua 2014. Meirinhos 2009[3]). Una expresión importante de este fenómeno son las redes sociales Twitter, Facebook, Instagram, Snapchat, Whatsapp, entre otras (Garcia 2006, Arguelles 2013[4]). En este contexto, las redes sociales que originalmente fueron pensadas como aplicaciones para distracción, comienzan a tener cada vez más presencia en otros ámbitos, ya que hoy es posible utilizar los datos que circulan a través de ellas para transformarlos en información de relevancia y utilidad sobre tendencias de consumo, pensamientos políticos, ideologías, preferencias y costumbres (Gil 2012, Lorca 2008, Miguel 2013[5]). Muchos de estos datos se encuentran disponibles para ser analizados y existen herramientas que pueden comprender el lenguaje natural, caracterizarlo y sistematizarlo (ronchetti 2016[6]).

En la comunidad científica existen antecedentes relacionados con este trabajo, donde se ha implementado análisis de opinión de los datos a partir de las redes sociales. Estos casos han sido utilizados en predicciones políticas y toma de decisiones para candidaturas políticas como las elecciones de Andalucía 2012 (Deltell 2013), campañas electorales en Estados Unidos 2012 e Italia 2012 (Ceron 2015) y Australia 2013 (Unankard 2014). En estos casos, las técnicas empleadas en el desarrollo de encuestas desde la base del análisis de datos de las redes sociales, representan una importante reducción de costos, tiempos, confección del cuestionario, diseño de muestras, testing del proceso, capacitación de los encuestadores y mejora de la gestión de la información, comparado con el enfoque tradicional de encuestas (Agudelo, Thayer-Hart 2010). Por tanto, para el procesamiento de los datos de Twitter en este trabajo se usó la base de datos columnar Vertica que provee un componente de software para el análisis de sentimientos llamando Pulse, siendo una iniciativa diferente a las propuestas tradicionales.

El desarrollo realizado propone una metodología de predicción sobre comportamientos sociales basada en análisis de sentimientos de tweets. El caso de estudio fue el Balotaje Presidencial de Argentina 2015 donde se utilizaron los datos que se generaron en Twitter, con el objetivo de predecir el resultado real de la elección.

Este trabajo se organiza de la siguiente forma: en la sección 2 se describe de manera general el caso de estudio seleccionado; la sección 3 presenta la configuración experimental empleada; la sección 4 muestra los resultados generales del trabajo. Finalmente, la sección 5 presenta las conclusiones del trabajo.

---

[2]Pla  http://journal.sepln.org/sepln/ojs/ojs/index.php/pln/article/view/5218/3022
[3] Cacheiro, M., Murua I., Gallego, D., (2015)Caracterización de las cibercomunidades de aprendizaje
Meirinhos, M., Osorio, A., (2009) Las comunidades virtuales de aprendizaje: el papel central de la colaboración
[4] García, Las redes sociales como herramientas para el aprendizaje colaborativo: una experiencia con Facebook
Arguelles, Las redes sociales y su aplicación en la educación. Revista Digital Universitaria
[5] Gil, , Redes sociales educativas como introducción a los entornos sociales de aprendizaje
Lorca, J., & Pujol, L. (2008). Redes Sociales: descripción del fenómeno, situación actual y perspectivas
Miguel ,V. (2013). Redes Sociales y Construcción del Conocimiento
[6] "Handshape recognition for Argentinian Sign Language using ... - SeDiCI." 2016. 12 May. 2016
<http://sedici.unlp.edu.ar/handle/10915/52376>

## 2. Descripción del caso de estudio

La República Argentina, país de 40.117.096 habitantes, según el censo de 2010, se encuentra en el hemisferio sur del continente Americano y está dividida en 24 provincias. En el marco del Balotaje de las elecciones presidenciales 2015 se condujo un estudio de corte transversal que incluyó 12.463.657 de tweets con una localización en el mismo país (ubicación geográfica del usuario de Twitter). Una muestra de los tweets, el analisis de sentimiento de cada tweets y la localizacion, se encuentran en Zenodo (https://zenodo.org/record/51496 ). Con esta información se puede validar el trabajo realizado.

Se utilizó análisis de sentimiento de textos, mediante el procesamiento de lenguaje natural desarrollando en un componente de software llamado Pulse, que incorpora Vertica (base de datos columnar y analítica). Para este fin se recolectaron los tweets en la base de datos referidos a las elecciones presidenciales desde el 5 al 25 de noviembre del 2015. Posteriormente, se calculó el análisis de sentimiento sólo de los tweets obtenidos hasta el 22 de noviembre de 2015 a las 17:59:59 de Argentina, fecha y hora de finalización del comicio electoral. Los atributos se categorizaron, dándole a los tweets los siguientes valores: positivo (1) para comentarios a favor del candidato, negativo (-1) en comentarios en contra del candidato y neutro (0) con comentarios que no están a favor de ninguno de los candidatos. Los candidatos eran Mauricio Macri y Daniel Scioli.

Finalmente se contrastó el cálculo de sentimiento positivo para cada uno de los candidatos con el resultado electoral oficial, para determinar relaciones, tanto a nivel nacional como por provincias, verificando los resultados por provincias al utilizar la ubicación geográfica del usuario de Twitter.

## 3. Configuración Experimental

### 3.1. Arquitectura

La Figura 1 describe la arquitectura de software empleada para la investigación, donde se utilizó la API de Twitter a través del Agente Apache Flume de Hadoop (AAF - Apache-flume-1.3.1). Este agente carga los tweets en la base de datos columnar Vertica (Vertica Analytic Database v7.1.2-0). Luego esta información es analizada con funciones analíticas de Vertica (Vertica-pulse-7.1.2-0) y la visualización de los resultados se realizan con la herramienta Tableau (Tableau 9.1). Finalmente, se utilizó el software Stata 14 MP para un análisis de correlación de los datos(STATA).

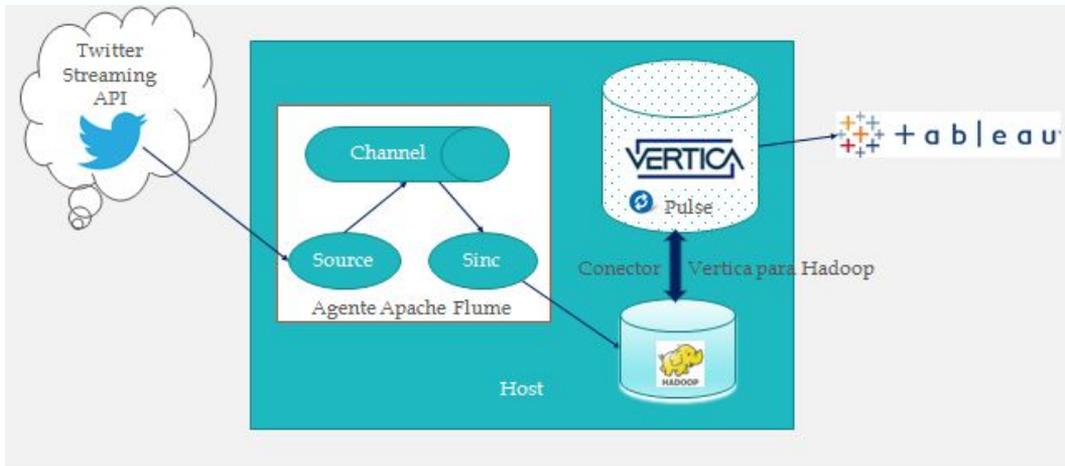

**Figura 1. Arquitectura de la solución**

## 3.2. Estructuras de Datos

La estructura de datos definida se compone de siete tablas y tres vistas, resumidas en la Tabla 1. El detalle de las tres principales tablas se puede observar en la Figura 2.

| Estructura de datos | Tipo | Descripción |
|---|---|---|
| tweets | tabla | Contiene los datos obtenidos de la API de Twitter. |
| tweet_sentiment | tabla | Contiene el cálculo de sentimiento para cada párrafo del tweet. |
| localidades | tabla | Contiene la relación entre el campo "user.location" de la tabla tweets con respecto a la provincia y país indicada por el usuario de Twitter. |
| white_list_es | tabla | Contiene los atributos sobre los cuales se analiza el sentimiento en los tweets mediante el lenguaje castellano. |
| normalization_es | tabla | Contiene sinónimos de los atributos definidos en la tabla "white_list_es". |
| pos_word_es | tabla | Palabras que se las consideran positivas. |
| neg_word_es | tabla | Palabras que se las consideran positivas. |
| v_tweets_score | vista | Calcula el score de cada tweet, sumando el resultado de cada párrafo. |
| v_tweets_candidatos | vista | Determina si el tweet nombra a uno de los dos candidatos. |
| v_sentiment_analisis | vista | Cuenta los tweets positivos, negativos o neutros por candidato, detallando el conteo por país y provincia. |

**Tabla 1. Estructura de datos empleada.**

De particular interés es el campo "text" de la tabla "tweets" ya que contiene los datos utilizados para calcular el sentimiento del tweet. Otro campo importante de esta tabla es "user.location" dado que se utilizará para relacionar el tweet con los resultados provinciales (ubicación geográfica). Debido a que un "tweet" puede estar formado por uno o más párrafos, la tabla "tweet_sentiment" contiene el cálculo de sentimiento asignado a cada párrafo del tweet en el campo "sentiment_score". Por ello, la vista "v_tweets_score" se utiliza para ponderar el sentimiento general del tweet a partir de la suma de sentimientos de sus párrafos: si la suma da por resultado un valor positivo, cero o negativo se le asignará el valor 1, 0 o -1 respectivamente al tweet en cuestión. La Figura 3 muestra el código de la consulta que genera la vista.

| Schema | Table | Column | Type | Size | Default | Not Null | Primary Key | Foreign Key |
|--------|-------|--------|------|------|---------|----------|-------------|-------------|
| elec | tweets | id | int | 8 | | t | t | |
| elec | tweets | created_at | timestamptz | 8 | | f | f | |
| elec | tweets | "user.name" | varchar(144) | 144 | | f | f | |
| elec | tweets | "user.screen_name" | varchar(144) | 144 | | f | f | |
| elec | tweets | "user.followers_count" | int | 8 | | f | f | |
| elec | tweets | "text" | varchar(500) | 500 | | f | f | |
| elec | tweets | "retweeted_status.retweet_count" | int | 8 | | f | f | |
| elec | tweets | "retweeted_status.id" | int | 8 | | f | f | |
| elec | tweets | "retweeted_status.favorite_count" | int | 8 | | f | f | |
| elec | tweets | "retweeted_status.text" | varchar(500) | 500 | | f | f | |
| elec | tweets | "user.location" | varchar(500) | 500 | | f | f | |
| elec | tweets | "coordinates.coordinates.0" | float | 8 | | f | f | |
| elec | tweets | "coordinates.coordinates.1" | float | 8 | | f | f | |
| elec | tweets | "entities.hashtags.0.text" | varchar(144) | 144 | | f | f | |
| elec | tweets | "entities.hashtags.1.text" | varchar(144) | 144 | | f | f | |
| elec | tweets | lang | varchar(5) | 5 | | f | f | |

(16 rows)

dbadmin=> \d elec.tweet_sentiment

List of Fields by Tables

| Schema | Table | Column | Type | Size | Default | Not Null | Primary Key | Foreign Key |
|--------|-------|--------|------|------|---------|----------|-------------|-------------|
| elec | tweet_sentiment | id | int | 8 | | f | | f | |
| elec | tweet_sentiment | "user.screen_name" | varchar(144) | 144 | | f | | f | |
| elec | tweet_sentiment | sentence | int | 8 | | f | | f | |
| elec | tweet_sentiment | attribute | varchar(500) | 500 | | f | | f | |
| elec | tweet_sentiment | sentiment_score | float | 8 | | f | | f | |

(5 rows)

dbadmin=> \d elec.localidades

List of Fields by Tables

| Schema | Table | Column | Type | Size | Default | Not Null | Primary Key | Foreign Key |
|--------|-------|--------|------|------|---------|----------|-------------|-------------|
| elec | localidades | "user.location" | varchar(500) | 500 | | t | t | |
| elec | localidades | province | varchar(50) | 50 | | f | f | |
| elec | localidades | country | varchar(50) | 50 | | f | f | |

**Figura 2. Descripción de Tablas**

Para determinar a qué candidato corresponde el sentimiento de cada tweet se utiliza la vista "v_tweets_candidatos". A través de esta vista se asigna al tweet los valores "Scioli", "Macri" "Scioli-Macri" o "Sin Candidato" en función del nombramiento del candidato, de los candidatos o de ninguno en el tweet. En la Figura 4 se muestra el código de la vista.

```
CREATE  VIEW elec.v_tweets_score AS
 SELECT a.id,
      CASE WHEN (a.sentiment_score = 0::float) THEN 0 WHEN (a.sentiment_score > 0::float) THEN 1
ELSE (-1) END AS sentiment_score
 FROM ( SELECT t.id,
      sum(s.sentiment_score) AS sentiment_score
FROM (elec.tweets t JOIN elec.tweet_sentiment s ON ((t.id = s.id)))
 GROUP BY t.id) a;
```

**Fig. 3. Código de la vista "v_tweets_score"**

Las tablas "white_list_es", "normalization_es", "pos_word_es" y "neg_word_es" forman parte de la herramienta Pulse y serán descritas posteriormente en la sección "Configuración de los diccionarios de Pulse".

```
CREATE  VIEW elec.v_tweets_candidatos AS
 SELECT tweets.id,
      CASE WHEN ((lower(tweets."text") ~~ '%macri%'::varchar(7)) AND (lower(tweets."text") ~~
'%scioli%'::varchar(8))) THEN 'Scioli-Macri'
::varchar(12) WHEN (lower(tweets."text") ~~ '%macri%'::varchar(7)) THEN 'Macri'::varchar(5) WHEN
(lower(tweets."text") ~~ '%scioli%'::varcha
r(8)) THEN 'Scioli'::varchar(6) ELSE 'Sin Candidato'::varchar(13) END AS candidato
 FROM elec.tweets;
```
**Fig. 4. Código de la vista "v_tweets_candidatos"**

La figura 5 describe la vista con los resultados por provincia, esta vista se utilizó para visualizar los datos en la herramienta Tableau

```
CREATE  VIEW elec.v_sentiment_analisis AS
 SELECT l.country,
      l.province,
      c.candidato,
      CASE s.sentiment_score WHEN 1 THEN 'Positivo'::varchar(8) WHEN (-1) THEN
'Negativo'::varchar(8) ELSE 'Neutro'::varchar(6) END
AS sentimiento,
      count(*) AS cantidad
 FROM (((elec.tweets t JOIN elec.v_tweets_candidatos c ON ((t.id = c.id))) JOIN elec.localidades l ON
((l."user.location" = t."user.lo
cation"))) JOIN elec.v_tweets_score s ON ((t.id = s.id)))
 GROUP BY l.country,
      l.province,
      c.candidato,
      s.sentiment_score;
```
**Fig. 5. Código de la vista "v_sentiment_analisis"**

## 3.3. Metodología

La metodología empleada consiste en tres procesos o tareas simultáneas que se repitieron hasta finalizar el balotaje:
- Obtención de la información
- Cálculo de sentimientos
- Visualización de resultado

Una vez escrutado el comicio se realizó el análisis confirmatorio estadístico, mediante análisis de correlación de las variables de tweets por cada candidato y los votos de la elección.

Obtención de la información

Dado que la captura de la información se realizó utilizando el Agente Apache Flume (AAF), la primera actividad fue definir la configuración del mismo. Esto significó indicarle a AAF cuáles hashtags, palabras clave y cuentas de twitter a seguir, para obtener toda la información de la temática de interés. Para determinar esos valores se realizó un estudio previo sobre tweets referidos a las elecciones, considerando las tendencias, trending topics y las cuentas de Twitter de los principales candidatos, medios digitales y programas de televisión políticos. Sin embargo, esta información fue actualizada periódicamente a medida que se avanzó con el trabajo. La Figura 6 muestra parte de la configuración del AAF al finalizar los comicios. En particular, la configuración de las propiedades "TwitterAgent.sources.Twitter.keywords" y "TwitterAgent.sources.Twitter.follow" de la configuración de AAF.

```
TwitterAgent.sources = Twitter
TwitterAgent.channels = MemChannel
TwitterAgent.sinks = Vertica

TwitterAgent.sources.Twitter.type = com.cloudera.flume.source.TwitterSource
TwitterAgent.sources.Twitter.channels = MemChannel
TwitterAgent.sources.Twitter.consumerKey = VeYpC9VSn6Aw60e5gUisfe5jp
TwitterAgent.sources.Twitter.consumerSecret = 2j11GKucWVukBXdZrtHLMmgVaOjQlAnFUx9yJh08T6JceW1oFV
TwitterAgent.sources.Twitter.accessToken = 546633630-B2jIJ1jtnbr11a59lPmtg0zRKglORQJ9Py46ZLEZ
TwitterAgent.sources.Twitter.accessTokenSecret = HXycStDgwHLQITxJsj5wp2v7rHCaRJHig3IYNQ1jqgHZL
TwitterAgent.sources.Twitter.keywords = Cambiemos,Cadena3Elecciones,Argentina,Elecciones2015,macri,fpv,Balotaje,ELECCIONES,Scioli,Ballotage,Cambiemos,yolovotoamm,peronismo,MacriPresidente,macripresidente,CambiemosConMacri,VamosConMacri,mm2015,MeHackearonLaCuenta,CampañaSucia,CampañaDeMiedo,YoVotoAScioli,MacriNosMiente,SiMacriGana,SiGanaMacri,ScioliPresidente,MIVICTORIA,Balotaje2015,argentinadebate,ganascioli,QueGaneScioli,Mesaza,GanoMacri,GanóMacri
TwitterAgent.sources.Twitter.follow = danielscioli,clarincom,pagina_12,infobae,C5N,6780ficial,mauriciomacri,TRIBUNAComar,todonoticias,lanacion,TV_PUBLICA
TwitterAgent.sources.Twitter.logging = false

# Define a logger sink that simply logs all events it receives
# and connect it to the other end of the same channel.
TwitterAgent.sinks.Vertica.channel = MemChannel
TwitterAgent.sinks.Vertica.type = org.apache.flume.sink.vertica.VerticaSink
TwitterAgent.sinks.Vertica.directory = ./../../files
TwitterAgent.sinks.Vertica.rollInterval = 10
TwitterAgent.sinks.Vertica.batchSize = 10000
TwitterAgent.sinks.Vertica.VerticaHost = 192.168.2.250
TwitterAgent.sinks.Vertica.port = 5433
TwitterAgent.sinks.Vertica.databaseName = verticadb
TwitterAgent.sinks.Vertica.username = dbadmin
TwitterAgent.sinks.Vertica.password = password
TwitterAgent.sinks.Vertica.tableName = elec.tweets
TwitterAgent.sinks.Vertica.parserName = TweetParser
TwitterAgent.sinks.Vertica.cleanup = true

TwitterAgent.channels.MemChannel.type = memory
TwitterAgent.channels.MemChannel.capacity = 10000
TwitterAgent.channels.MemChannel.transactionCapacity = 500
[dbadmin@localhost conf]$
```

**Figura 6. Configuración del Agente Apache Flume**

Por otro lado, entre los datos que entrega la API de Twitter, se obtiene el dato de localidad geográfica en el campo "user.location". Es importante destacar que esta información puede ser manipulada por el usuario de Twitter y puede no corresponder a una localidad correcta de la República Argentina, por lo cual a los efectos de relacionar las localidades con las provincias argentinas se generó una tabla denominada "localidades". Esta tabla se usó para relacionar "user.location" con la provincia y el país. A medida que se recibían los tweets se fueron agregando registros a la tabla localidades para vincular el tweet a la provincia. La Figura 6 muestra un ejemplo de la instrucción empleada para la normalización de los datos.

```
insert into elec.localidades ("user.location")
select "user.location"
from elec.tweets
where "user.location"
not in (select "user.location" from elec.localidades)
group by "user.location"
having "user.location" is not null;

update elec.localidades set province='Capital Federal', country='Argentina'
where lower("user.location") like '%caba%' and province is null;
commit;
```

**Figura 7. Ejemplo de instrucción para la normalización de los datos de ubicación del tweet.**

Cálculo de sentimientos

El análisis de sentimiento se realizó utilizando la herramienta Pulse de la base de datos analítica Vertica.

Para esto es necesario configurar la herramienta insertando valores en las tablas "white_list_es", "normalization_es", "pos_word_es" y "neg_word_es" descritas en la Tabla 1. Los valores de esta tabla serán utilizados luego por la función de cálculo de sentimiento para clasificar los tweets en positivo (1), negativo (-1) o neutro (0). La Figura 8 muestra la configuración de atributos para este trabajo.

```
dbadmin=> select * from pulse.white_list_es;
 standard
----------
 macri
 scioli
(2 rows)
```

**Figura 8. Atributos utilizados por pulse para el cálculo de sentimiento.**

Dado que en el texto pueden aparecer palabras o frases que identifican a los atributos sin nombrarlos, la tabla "normalization_es" se puede cargar con palabras que sean sinónimos de los atributos. La Figura 9 muestra una consulta sobre esta tabla. Por ejemplo, para indicar que la ocurrencia de palabra "#Cambiemos" en un tweet es un sinónimo de "macri", se debe insertar el registro correspondiente en esta tabla.

```
dbadmin=> select * from pulse.normalization_es;
 standard_base | standard_synonym
---------------+--------------------
 macri         | #Cambiemos
 macri         | #CambiemosConMacri
 ...
 scioli        | #CaravanaNaranja
 scioli        | #ElDebateEnTVP
 scioli        | #FPV
(52 rows)
```

**Figura 9. Ejemplo de sinónimos configurados en la tabla "normalization_es".**

Para determinar si el sentimiento del texto analizado es positivo o negativo, se deben cargar las tablas "pos_words_es" y "neg_words_es" con las palabras que se consideren positivas o negativas respectivamente. Pulse considerará las palabras de estas tablas influyentes en un sentido u otro sobre los textos a analizar. La Figura 10 y la Figura 11 muestran ejemplos de consultas sobre esas tablas.

```
dbadmin=> select * from pulse.pos_words_es;
         standard
-------------------------------
 678ConLaGenteAdentro
 Cambiemos
 CambiemosConMacri
 CaravanaNaranja
 …
 yolovotoamm
```

(38 rows)

**Figura 10. Consulta sobre la tabla "pos_word_es".**

```
dbadmin=> select * from pulse.neg_words_es;
        standard
--------------------------------
 2001
 90
 Anibal fernandez
 Aníbal
 CampañaDeMiedo
...
 aumentará los impuestos
(134 rows)
```

**Figura 11. Consulta sobre la tabla "neg_word_es".**

Finalmente, el cálculo de sentimiento se realiza mediante la función de Pulse "SentimentAnalysis()", y sus resultados se insertan en la tabla tweet_sentiment de acuerdo con la instrucción "insert" descrita en la Figura 12.

```
insert into elec.tweet_sentiment
(select id,
"user.screen_name",
SentimentAnalysis(text using parameters filterlinks=true,filterusermentions=false,
filterhashtags=false)
over (partition by id, "user.screen_name", text)
from elec.tweets where lang='es'
and created_at between to_timestamp ('2015-11-20 00:00:00','YYYY-MM-DD HH:MI:SS') and
to_timestamp ('2015-11-20 23:59:59','YYYY-MM-DD
HH:MI:SS'));
commit;
```

**Figura 12. Instrucción utilizada para realizar el cálculo de sentimiento sobre los tweets.**

Una vez calculado el sentimiento sobre los tweets, se verifica manualmente los resultados. En caso de encontrar errores o incoherencias se ajustan las tablas de Pulse descritas anteriormente, se trunca la tabla tweet_sentiment y se repite el cálculo. Para acelerar el proceso de ajuste de los diccionarios se generó una tabla temporal con todas las palabras utilizadas en los tweets ordenadas por cantidad de ocurrencias en forma descendente y luego se agregaron a los diccionarios las palabras de mayor frecuencia identificándose como positivas o negativas. En la Figura 13 se describe el proceso realizado. Al igual que con la configuración de AAF, este proceso se repitió hasta finalizados los comicios.

```
dbadmin=> create table elec.words(words varchar(500), input varchar(500));

dbadmin=> insert into elec.words SELECT TxtIndex.StringTokenizerDelim(text,' ') over()
dbadmin=> from elec.tweets;
```

```
dbadmin=> commit;
dbadmin=> select words, count(*) from elec.words group by words order by 2 desc;

     words          | count
------------------------+---------
 Macri              | 1899614
 Argentina          | 1416388
 Scioli             | 915487
 elecciones         | 565196
 #MacriPresidente   | 484177
 @mauriciomacri:    | 388416
 @mauriciomacri     | 357536
 Mauricio           | 333562
 #ArgentinaDebate   | 326241
 todos              | 266850
 #Argentina         | 244612
 macri              | 244332
 presidente         | 242502
 #Elecciones2015    | 240109
 Venezuela          | 229394
 #ScioliPresidente  | 227044
 porque             | 221806
 #Cambiemos         | 220883
 argentina          | 218805
 MACRI              | 212296
 Macri,             | 207646
```

**Figura 13. Proceso de ajuste de configuración.**

<u>Visualización</u>

Para la visualización de los datos se utilizó la herramienta Tableau. Esta herramienta permite de manera sencilla conectarse con Vertica y construir gráficos, tablas y mapas, en este caso referidos a los resultados nacionales y por provincias. Para mostrar la información en Tableau se utilizó la vista "v_sentiment_analisis" descrita en la Tabla 1 y Listado 3, que unifica los cálculos de sentimiento por cada tweet y los asigna a un candidato y a una provincia. En la siguiente sección se muestran los resultados finales obtenidos.

<u>Análisis confirmatorio</u>

Posteriormente se realizó un análisis de correlación para verificar la asociación entre los tweets positivos de cada candidato y los votos obtenidos en el balotaje electoral presidencial 2015.

**4. Resultados**

En esta sección se presentan los resultados finales de aplicar la metodología descrita. Para una mejor interpretación se presentan los resultados a nivel nacional y luego a nivel provincial.

**4.1 Resultados nacionales**

La Figura 14 muestra la consulta utilizada al cierre de los comicios. Los resultados se calcularon sobre la base de "tweets positivos" sobre cada candidato. En total, se pudieron clasificar 579.495 tweets como positivos entre ambos candidatos, de los cuales el 51.55% fue a favor del candidato Macri y el 48.44% a favor del candidato Scioli.

```
dbadmin=>
dbadmin=> select add_vertica_options('OPT','ENABLE_WITH_CLAUSE_MATERIALIZATION');
 add_vertica_options
--------------------
 Options Set
(1 row)

dbadmin=> with res as (select c.candidato,
dbadmin(>                 case s.sentiment_score when 1 then 'Positivo' when -1 then 'Negativo' else
'Neutro' end sentimiento,
dbadmin(>                 count(*) cantidad
dbadmin(>             from elec.tweets t inner join elec.v_tweets_candidatos c on (t.id=c.id)
dbadmin(>             inner join elec.localidades l on (l."user.location"=t."user.location")
dbadmin(>             inner join elec.v_tweets_score s on (t.id=s.id)
dbadmin(>             where c.candidato in ('Scioli','Macri') and s.sentiment_score=1
dbadmin(>             group by c.candidato, s.sentiment_score)
dbadmin->
dbadmin-> select candidato, sentimiento, cantidad , cantidad*100/(select sum(cantidad) from res)
porcentaje
dbadmin-> from res;

 candidato | sentimiento | cantidad |      porcentaje
-----------+-------------+----------+-----------------------
 Macri     | Positivo    |   298746 | 51.552817539409313281
 Scioli    | Positivo    |   280749 | 48.447182460590686719
(2 rows)
```

**Figura 14. Consulta y resultados obtenidos al finalizar los comicios.**

Del total de 12.463.657 de tweets, se utilizaron para el cálculo de los porcentajes 579.495 ya que solo se tuvo en cuenta para este estudio los tweets catalogados como positivos donde el texto del tweet nombro solo a uno de los dos candidatos, descartando en este caso los cálculos de sentimiento negativos y neutros. Otra de las consideraciones para el cálculo del resultado fue que el cálculo de sentimiento se realizó solo hasta la hora del cierre del comicio, descartando los tweets posteriores.

Por otro lado la Figura 15 muestran los resultados oficiales publicados por la comisión electoral el día 25 de noviembre de 2015, tres días después de las elecciones.

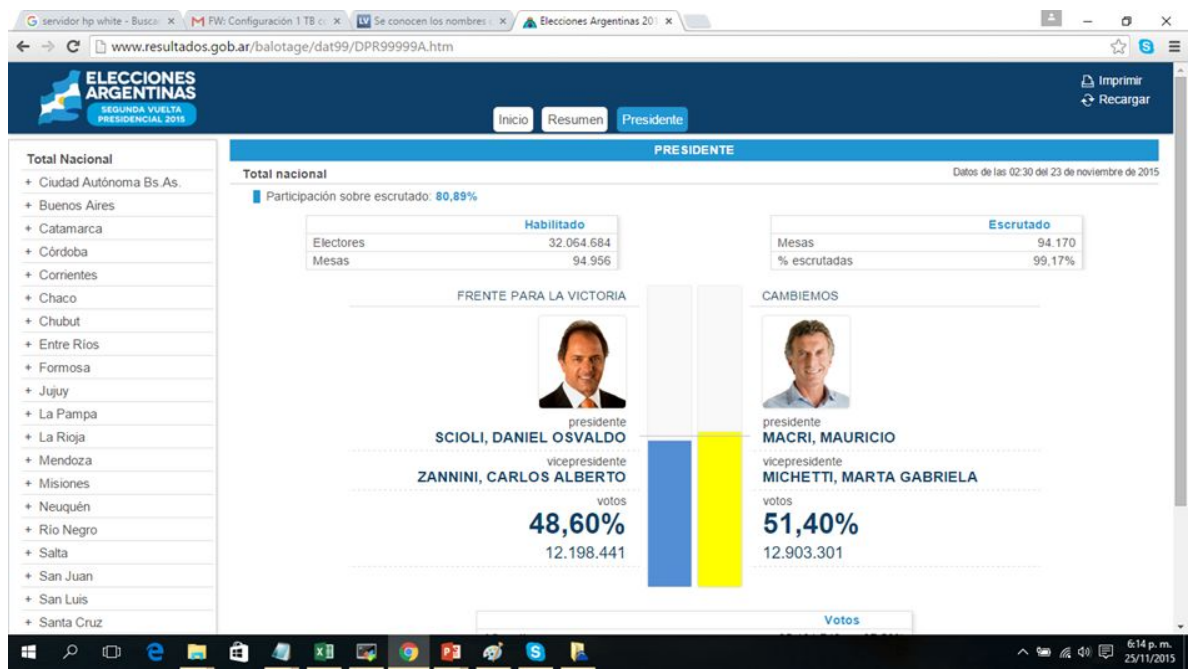

**Figura 15. Resultados publicados por la comisión electoral**

Es de interés destacar que el margen de error de los resultados calculados al cierre del balotaje en este trabajo con respecto a los resultados oficiales fue aproximadamente de 0.16 puntos porcentuales. Por último, la Figura 16 muestra las predicciones realizadas por los encuestadores de opinión tradicionales. En este caso, todos ellos tuvieron un margen de error a nivel nacional mayor que el de nuestro estudio.

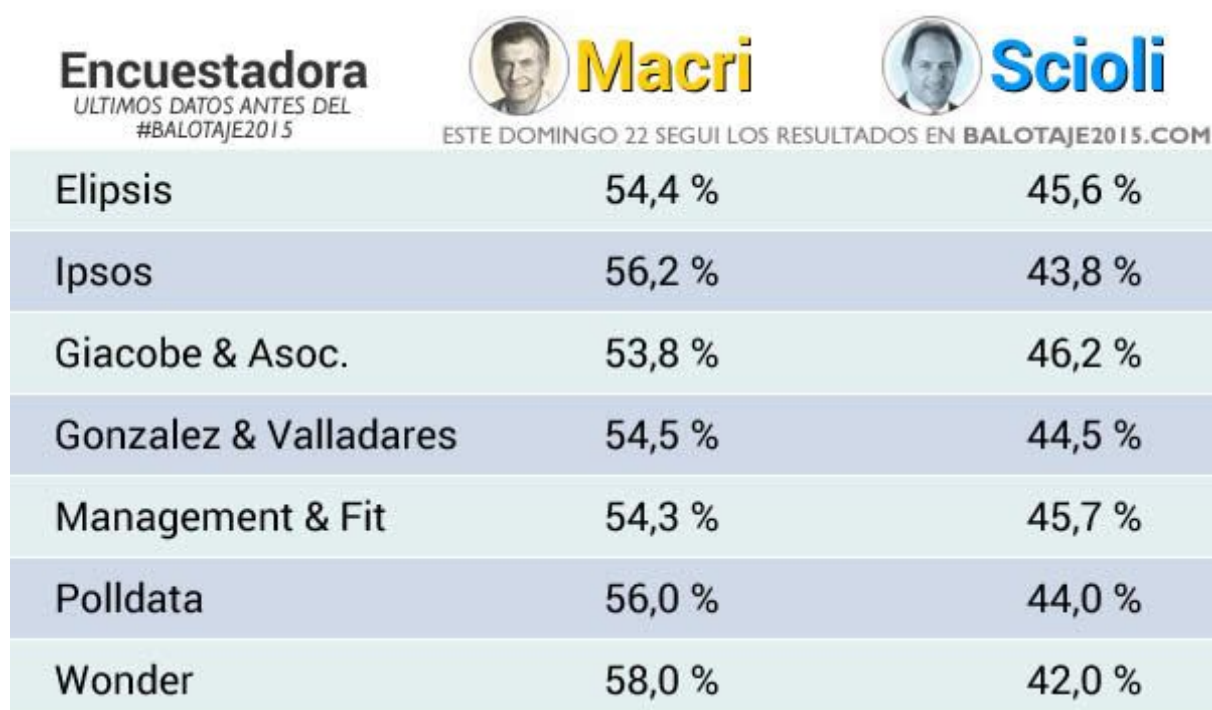

**Figura 16. Encuestas de intención de voto. Fuente: http://balotaje2015.com/encuestas/**

### 4.2 Tweets positivos por provincia por cada candidato

Un segundo análisis buscando relacionar los resultados de acuerdo con la ubicación del tweet sobre las provincias que componen el territorio argentino, determinó que en 16 provincias de 24, la predicción realizada coincidió con los resultados oficiales, La Figura 17 muestra la descomposición por provincias realizada con la herramienta Tableau, mientras que la Figura 18 muestra los resultados oficiales publicados por la comisión electoral.

Es importante aclarar que del total de 579.495 tweets positivos para un candidato u otro, solo se pudieron relacionar con las provincias un 55.02%, es decir, 318.857 tweets. Esto se debe a que en el proceso de normalización geográfica no se pudieron relacionar todos los datos de localización "tweets.user.location" con la provincia.

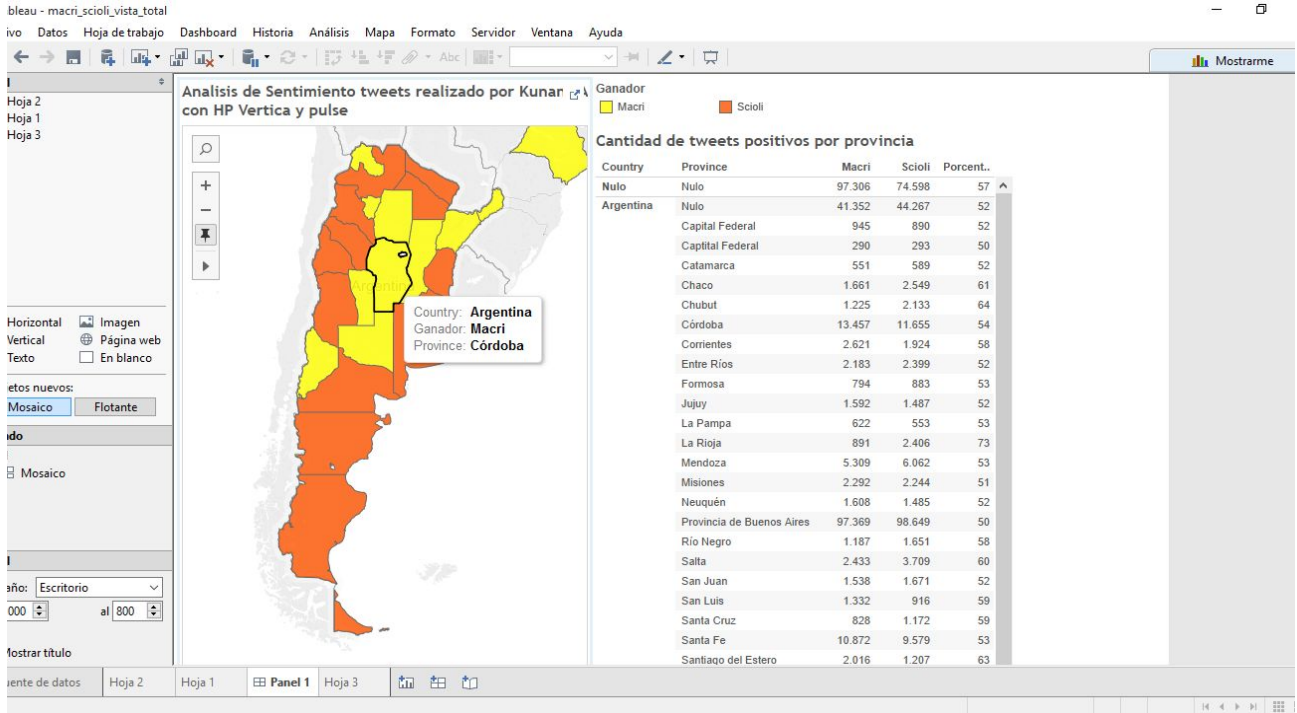

**Figura 17. Resultados de análisis de sentimiento discriminados por provincias**

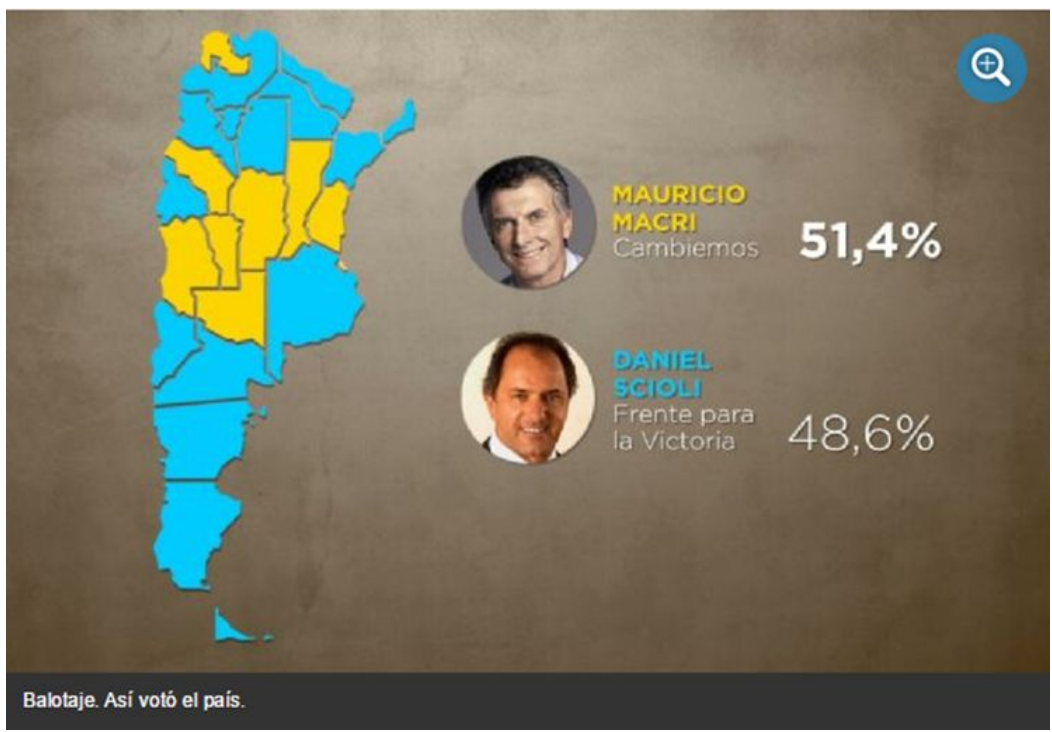

**Figura 18. Resultados oficiales discriminados por provincias**

En el Cuadro 2 se muestran los resultados del análisis de sentimiento positivo por provincia, y la relación entre la cantidad de tweets analizados (tweets positivos), los votos escrutados (votos) y la población Argentina en el censo 2010 (población). Como se puede observar el 44,98% de los datos, "Sin Provincia", no pudieron normalizarse.

Si bien en 16 de las 24 provincias el resultado del análisis de tweets coincide con el resultado de la elección, los porcentajes de 10 de las provincias, Córdoba, Salta, Capital Federal, Tierra del Fuego, Mendoza, Tucumán, Entre Ríos, Misiones, La Rioja y Santiago del Estero, muestran diferencias de más de 5 puntos entre el resultado de análisis de sentimiento y la elección ("% tweets por prov" y "% votos por prov"). En la mayoría de las provincias se nota una proporcionalidad de la cantidad de tweets recolectados con respecto al porcentaje total de la población y de votantes, salvo en Capital Federal, donde el porcentaje del total de los votos es del 7,72% y el porcentaje de tweets respecto al total es de tan solo el 0,42%.

| country | province | tweets positivos | | | votos | | | población | % tweets por prov | | % votos por prov | | relacion muestra | | |
|---|---|---|---|---|---|---|---|---|---|---|---|---|---|---|---|
| | | scioli | macri | tweets | scioli | macri | total | | %tw.scioli | %tw.macri | %v.scioli | %v.macri | %tweets | %votos | %pob |
| Argentina | Provincia de Buenos Aires | 98649 | 97369 | 196018 | 4.833.680 | 4.626.326 | 9.460.006 | 15.625.084 | 50,33 | 49,67 | 51,10 | 48,90 | 33,83 | 37,69 | 38,95 |
| Sin País | Sin Provincia | 74598 | 97306 | 171904 | | | | | 43,40 | 56,60 | | | 29,66 | 0,00 | 0,00 |
| Argentina | Sin Provincia | 44267 | 41352 | 85619 | | | | | 51,70 | 48,30 | | | 14,77 | 0,00 | 0,00 |
| Argentina | Córdoba | 11655 | 13457 | 25112 | 613.611 | 1.540.018 | 2.153.629 | 3.308.876 | 46,41 | 53,59 | 28,49 | 71,51 | 4,33 | 8,58 | 8,25 |
| Argentina | Santa Fe | 9579 | 10872 | 20451 | 902.910 | 1.136.478 | 2.039.388 | 3.194.537 | 46,84 | 53,16 | 44,27 | 55,73 | 3,53 | 8,12 | 7,96 |
| Argentina | Mendoza | 6062 | 5309 | 11371 | 461.812 | 625.121 | 1.086.933 | 1.738.929 | 53,31 | 46,69 | 42,49 | 57,51 | 1,96 | 4,33 | 4,33 |
| Argentina | Tucumán | 3508 | 3924 | 7432 | 562.602 | 397.559 | 960.161 | 1.448.188 | 47,20 | 52,80 | 58,59 | 41,41 | 1,28 | 3,83 | 3,61 |
| Argentina | Salta | 3709 | 2433 | 6142 | 373.800 | 314.944 | 688.744 | 1.214.441 | 60,39 | 39,61 | 54,27 | 45,73 | 1,06 | 2,74 | 3,03 |
| Argentina | Entre Ríos | 2399 | 2183 | 4582 | 386.874 | 450.998 | 837.872 | 1.235.994 | 52,36 | 47,64 | 46,17 | 53,83 | 0,79 | 3,34 | 3,08 |
| Argentina | Corrientes | 1924 | 2621 | 4545 | 353.909 | 285.351 | 639.327 | 992.595 | 42,33 | 57,67 | 55,37 | 44,63 | 0,78 | 2,55 | 2,47 |
| Argentina | Misiones | 2244 | 2292 | 4536 | 384.954 | 278.308 | 663.262 | 1.101.593 | 49,47 | 50,53 | 58,04 | 41,96 | 0,78 | 2,64 | 2,75 |
| Argentina | Chaco | 2549 | 1661 | 4210 | 401.056 | 276.057 | 677.113 | 1.055.259 | 60,55 | 39,45 | 59,23 | 40,77 | 0,73 | 2,70 | 2,63 |
| Argentina | Chubut | 2133 | 1225 | 3358 | 185.142 | 129.515 | 314.657 | 509.108 | 63,52 | 36,48 | 58,84 | 41,16 | 0,58 | 1,25 | 1,27 |
| Argentina | La Rioja | 2406 | 891 | 3297 | 87.793 | 114.092 | 201.885 | 333.642 | 72,98 | 27,02 | 43,49 | 56,51 | 0,57 | 0,80 | 0,83 |
| Argentina | Santiago del Estero | 1207 | 2016 | 3223 | 399.189 | 154.742 | 553.931 | 874.006 | 37,45 | 62,55 | 72,06 | 27,94 | 0,56 | 2,21 | 2,18 |
| Argentina | San Juan | 1671 | 1538 | 3209 | 256.875 | 172.948 | 429.823 | 681.055 | 52,07 | 47,93 | 59,76 | 40,24 | 0,55 | 1,71 | 1,70 |
| Argentina | Neuquén | 1485 | 1608 | 3093 | 198.873 | 177.350 | 376.223 | 551.266 | 48,01 | 51,99 | 52,86 | 47,14 | 0,53 | 1,50 | 1,37 |
| Argentina | Jujuy | 1487 | 1592 | 3079 | 186.202 | 209.342 | 395.544 | 673.307 | 48,29 | 51,71 | 47,07 | 52,93 | 0,53 | 1,58 | 1,68 |
| Argentina | Río Negro | 1651 | 1187 | 2838 | 250.143 | 147.876 | 398.019 | 638.645 | 58,17 | 41,83 | 62,85 | 37,15 | 0,49 | 1,59 | 1,59 |
| Argentina | Capital Federal | 1183 | 1235 | 2418 | 682.354 | 1.255.099 | 1.937.453 | 2.890.151 | 48,92 | 51,08 | 35,22 | 64,78 | 0,42 | 7,72 | 7,20 |
| Argentina | San Luis | 916 | 1332 | 2248 | 98.979 | 177.183 | 276.162 | 432.310 | 40,75 | 59,25 | 35,84 | 64,16 | 0,39 | 1,10 | 1,08 |
| Argentina | Santa Cruz | 1172 | 828 | 2000 | 101.270 | 72.239 | 173.509 | 273.964 | 58,60 | 41,40 | 58,37 | 41,63 | 0,35 | 0,69 | 0,68 |
| Argentina | Tierra del Fuego | 1126 | 577 | 1703 | 54.529 | 38.332 | 92.861 | 127.205 | 66,12 | 33,88 | 58,72 | 41,28 | 0,29 | 0,37 | 0,32 |
| Argentina | Formosa | 883 | 794 | 1677 | 205.636 | 115.746 | 321.382 | 530.162 | 52,65 | 47,35 | 63,96 | 36,04 | 0,29 | 1,28 | 1,32 |
| Argentina | La Pampa | 553 | 622 | 1175 | 104.013 | 107.888 | 211.901 | 318.951 | 47,06 | 52,94 | 49,09 | 50,91 | 0,20 | 0,84 | 0,80 |
| Argentina | Catamarca | 589 | 551 | 1140 | 112.368 | 99.789 | 212.157 | 367.828 | 51,67 | 48,33 | 52,96 | 47,04 | 0,20 | 0,85 | 0,92 |
| España | Madrid | 375 | 468 | 843 | | | 0 | | 44,48 | 55,52 | | | 0,15 | 0,00 | 0,00 |
| Uruguay | Montevideo | 411 | 377 | 788 | | | 0 | | 52,16 | 47,84 | | | 0,14 | 0,00 | 0,00 |
| Colombia | Bogotá | 131 | 498 | 629 | | | 0 | | 20,83 | 79,17 | | | 0,11 | 0,00 | 0,00 |
| Venezuela | Caracas | 156 | 440 | 596 | | | 0 | | 26,17 | 73,83 | | | 0,10 | 0,00 | 0,00 |
| Venezuela | Sin Provincia | 36 | 116 | 152 | | | 0 | | 23,68 | 76,32 | | | 0,03 | 0,00 | 0,00 |
| Brasil | Río de Janeiro | 26 | 52 | 78 | | | 0 | | 33,33 | 66,67 | | | 0,01 | 0,00 | 0,00 |
| Brasil | São Paulo | 6 | 20 | 26 | | | 0 | | 23,08 | 76,92 | | | 0,00 | 0,00 | 0,00 |
| Uruguay | Sin Provincia | 1 | 0 | 1 | | | 0 | | 100,00 | 0,00 | | | 0,00 | 0,00 | 0,00 |
| Colombia | Sin Provincia | 1 | 0 | 1 | | | 0 | | 100,00 | 0,00 | | | 0,00 | 0,00 | 0,00 |
| España | Sin Provincia | 1 | 0 | 1 | | | 0 | | 100,00 | 0,00 | | | 0,00 | 0,00 | 0,00 |
| | | | | | | | 0 | | | | | | | | |
| TOTALES | | 280749 | 298746 | 579495 | 12.198.441 | 12.903.301 | 25.101.742 | 40.117.096 | 48,44718 | 51,55282 | 48,60 | 51,40 | 100 | 100,00 | 100 |

**Cuadro 2 . Relación entre la población y los tweets analizados por provincia**

### 4.3 Análisis de Correlación entre las variables

El análisis de correlación verificó una asociación significativa entre los tweets positivos de cada candidato y los votos obtenidos en el balotaje electoral (resultados oficiales). Para los candidatos Macri y Scioli los coeficiente de correlación entre los tweets positivos y sus votos en el escrutinio fueron de 0.95 (p = 0,0001) y 0,98 (p = 0,00001), respectivamente.

## 5. Conclusiones

Mediante el estudio realizado se ha logrado predecir el resultado electoral oficial a nivel nacional con una diferencia del 0,16%. Sin embargo, a nivel de agregación provincial, este porcentaje se incrementa, probablemente por la incapacidad de obtener el dato de localidad en forma confiable desde la API de Twitter, la cual impide establecer la relación entre la localización geográfica del usuario de twitter y la provincia.

Así, las herramientas utilizadas en este trabajo (Big Data, el análisis de sentimiento, datos de redes sociales), se encuentran en evolución permanente, por lo que continuar esta línea de investigación sería de gran trascendencia e importancia para diferentes campos laborales y ciencias, que pueden verse favorecidos por su utilización. Son necesarias futuras investigaciones para optimizar, actualizar y sistematizar los mecanismos utilizados en la presente investigación.

## 6. Referencias

**TIPS**
- HPE Vertica my.vertica.com
- http://www.nosql.es/blog/nosql/mapreduce.html
- PostgreSQL – http://www.postgresql.org/
- MongoDB – http://www.mongodb.org/
- Cassandra – http://cassandra.apache.org/
- Redis – http://redis.io/
- Neo4J – http://neo4j.com/
- Análislis IoT
- http://www.jcc.info.unlp.edu.ar/jcc2016/wordpress/index.php/envio-de-articulos/